\documentclass[12pt,dvips]{article}

\makeatletter
\renewcommand{\section}{\@startsection{section}{1}{0in}
	{0.4\baselineskip}{0.1\baselineskip}{\Large\bf}}
\renewcommand{\subsection}{\@startsection{subsection}{2}{0in}
	{0.25\baselineskip}{-\baselineskip}{\large\bf}}
\renewcommand{\subsubsection}{\@startsection{subsubsection}{3}{0in}
	{0.1\baselineskip}{-\baselineskip}{\normalsize\bf}}
\makeatother
\begin{document}

%
\makeatletter\newcommand{\ps@icrc}{
\renewcommand{\@oddhead}{\slshape{ICRC 2001 Posters
(updated comment)}\hfil}}
\makeatother\thispagestyle{icrc}
%
%

\begin{center}
%
~ 

{\LARGE \bf Deformed Lorentz Symmetry\\ 
and High-Energy Astrophysics (III)}
\end{center}

\begin{center}
%
%
~ 

{\bf Luis Gonzalez-Mestres$^{1,2}$}\\

~ 

{\it $^{1}$ L.A.P.P., CNRS-IN2P3, B.P. 110, 74941 Annecy-le-Vieux Cedex, France \\
$^{2}$ L.I.G.A., CNRS-ENS, 45 rue d'Ulm, 75230 Paris Cedex 05, France }
\end{center}

~ 

\begin{center}
{\large \bf Abstract\\}
\end{center}
\vspace{-0.5ex}
%
%
Lorentz symmetry violation (LSV) 
can be generated at the Planck scale, or at some
other fundamental length scale, and
naturally preserve Lorentz symmetry
as a low-energy limit (deformed Lorentz symmetry, DLS). 
DLS can
have important implications for ultra-high energy cosmic-ray physics
(see papers physics/0003080 - hereafter referred to as I - , astro-ph/0011181 and astro-ph/0011182, and references quoted in these papers). A crucial
question is how DLS can be extended to a deformed Poincar\'e symmetry (DPS), and what can be the dynamical origin of this phenomenon. In a recent paper ( hep-th/0208064 , hereafter referred to as II), we started a discussion of proposals to identify DPS with a symmetry incorporating the Planck scale (like doubly special relativity, DSR) and suggested new ways in similar directions. Implications for models of quadratically deformed relativistic kinematics (QDRK) and linearly deformed relativistic kinematics (LDRK) were also discussed. We pursue here our study of these basic problems, focusing on the possibility to relate deformed relativistic kinematics (DRK) to new space-time dimensions and compare our QDRK model, in the form proposed since 1997, which the Kirzhnits-Chechin (KCh) and Sato-Tati (ST) models. It is pointed out that, although the KCh model does not seem to work such as it was formulated, our more recent proposals can be related to suitable extensions of this model generalizing the Finsler algebras (even to situations where a preferred physical inertial frame exists) and using the Magueijo-Smolin transformation as a technical tool.

~ 

Together with II , this paper updates and further develops some aspects
of contributions to the ICRC 2001 Conference, Hamburg August 2001 (Posters: 0G 092 , OG 093 and HE 313).
~
~
\vspace{1ex}

~

%
%
~ 

\section{Introduction}
\label{introduction.sec}

~
~ 

The idea of doubly special relativity, DSR (Amelino-Camelia, Benedetti and D'Andrea, 2002; Amelino-Camelia, 2001 and 2002; Bruno, Amelino-Camelia and Kowalski-Gilman, 2001), and its analysis in (Lukierski and Nowicki 2002) and other recent papers, have raised the question of how far is it possible to associate the usual phenomenological models of deformed relativistic kinematics (DRK, see for instance: Gonzalez-Mestres, 1997a to 1997e, 1998, 2000a to 2000c) to more conventional formulations preserving Lorentz symmetry in specific reference systems. Magueijo and Smolin, 2001 and 2002, and the present author (Gonzalez-Mestres, 2002) have recently further explored the use of new transformations to relate a DRK universe to a formally conventional, relativistic one. As already emphasized in II, the existence in each physical inertial frame of such transformations for single-particle states does not by itself imply that they have the same form in all physical inertial frames as it is often postulated. Thus, a vacuum rest frame may still exist and DRK is not, in general, physically equivalent to DSR. The laws of physics are not necessarily the same in all physical inertial systems. But the Magueijo-Smolin transformation provides in any case a very useful tool to manipulate equations and discuss concepts in a DRK scenario.

~

It must be emphasized that our definition of deformed relativistic kinematics is phenomenological. By DRK and DLS or DPS we mean any pattern which tends to special relativity in the small-momentum scale but slightly deviates from it at finite momentum. Usually, the deviation becomes more important as energy increases, but remains small above some critical energy scale. The main phenomenological interest of a close study of DLS and DRK is that it can lead to observable physical predictions at energies well below the critical scale.

~

As in (Gonzalez-Mestres, 2002), we assume in this paper that a reference dynamical system (the "symmetric" system) exists with a standard Poincar\'e group. But the real physical reference system is a different one, where the energy and momentum conserved in the physical space-time do not correspond to those of the standard underlying Poincar\'e group. In the "physical" reference system, energy and momentum conservation is preserved. Energy is additive, as well as momentum, for a system of several non-interacting bodies. We also assume the existence of a physical vacuum rest frame, where everything is formulated in spite of the formal underlying Lorentz invariance in another reference system. Following Poincar\'e (1895, 1901, 1905), we interpret special relativity as the impossibility to identify the vacuum rest frame (VRF) by available physical measurements. As stressed in II, the question of energy and momentum conservation in the "symmetric" reference system (SRS) must be discussed keeping in mind that: a) the space-time is not the same as the "physical" space-time and does not commute with it; b) since the "physical" reference system (PRS) is the experimentally correct one (otherwise, everything would be simpler in the SRS), a fundamental instability must be at work in the SRS. This point will be discussed in a forthcoming paper.

~

The aim of this paper is to complete the discussion started in II about the $U$ operators introduced by Magueijo and Smolin and the potentialities of this technique applied to our phenomenological DRK approach, and to compare the present models with previously existing ones.

~

As in II, we assume that the transition between the ideal "symmetric" reference system where Lorentz symmetry is preserved and the (experimentally natural) "physical" one is driven by a singular non-unitary operator which, for a single-particle state, has the form:

\equation
U ~ (p_0 ~ , ~ p_i ~, ~ T ~ , ~ T_0 ~ , ~ a) 
\endequation
\noindent
and transforms a physical state $|\psi> ~ (physical)$ into an object in the SRS, $|\psi> ~ (symmetric)$ :

\equation
|\psi> ~ (symmetric) ~ = ~ U ~ |\psi> ~ (physical)
\endequation
\noindent
where $p_0$ is the energy, $p_i$ stands for momentum in the three space directions, $T$ is an intrinsic time scale associated to each particle, $T_0$ is a universal time scale, and $a$ is the fundamental length scale (e.g. the Planck length). The speed of light $c$ and the Planck constant are taken to be equal to 1 . We ignore the problems raised by singularities but, as shown in II, it is likely that in most cases an equivalent unitary operator $U'$ can be found. If a "physical" vacuum rest frame (PVRF) exists, $U$ may also depend on a parameter related to a relative speed with respect to the equivalent SRS (the "symmetric" vacuum rest frame, SVRF), in which case the laws of Physics would not be the same in all physical inertial frames. The properties of space-time dimensions are also a crucial question and will be discussed in detail in paper IV of this series .

~

Following II, we write for the energy-momentum quadrivector operator:

\equation
p_{\alpha} ~ (symmetric) ~ = ~ U ~ p_{\alpha} ~ (physical) ~ U^{-1}
\endequation
\noindent
where "physical" stands for the measured energy and momentum, and "symmetric" for energy and momentum in the ideal symmetric frame (the SRS). The index $\alpha ~ = ~ 0 ~ , ~ 1 ~ , ~ 2, ~ 3$ corresponds to energy and to the three momentum coordinates, with the metrics (+, -, -, -). In what follows, we refer to space indices as $i$ or $j$ = 1 , 2, 3 .

~
~ 

\section{Deformations and new space-time dimensions}
\label{deformation.sec}

~

In the "symmetric" reference system, we add, like in the Kaluza-Klein theory, an extra dimension described by the new momentum variable $\sigma $ and, to illustrate our basic idea, write the equation of motion for free particles:

\equation
p_{0}^2 ~ (symmetric) ~ = ~ p^2 ~ (symmetric) ~ + ~ \sigma ^2 ~ (symmetric)
\endequation
\noindent
where $p^2 ~ = ~ \Sigma _{i=1} ^3 ~ p_{i}^2$ .

~

We then take $U$ to be:

\equation
U (p_0 ~ , ~ p_i ~,~ \sigma  ~ , ~ T ~ , ~ T_0 ~ , ~ a) ~ = exp ~ [ln ~ (1 ~ - ~ b ~ p^2 ~ \sigma ^{-1}) ~ ~D_{space-2 , \sigma}]
\endequation
\noindent
where $b$ is a constant and $D_{space-2 , \sigma} ~ = p^2 ~ \partial /\partial p^2 ~ + ~ \sigma ~ \partial /\partial \sigma $ a modified dilatation generator for the space and $\sigma$ momentum dimensions. The symbol $ln$ stands for neperian logarithm. In the limit $b ~ \rightarrow ~ 0$ , one has $U ~ \rightarrow ~ 1$ . From (4) and (5), we get:

\equation
p_0 ~ (symmetric) ~ = ~ p_0 ~ (physical)
\endequation
\equation
p^2 ~ (symmetric) ~ = ~ p^2 ~ (physical) [1 ~ - ~ b ~ p^2 (physical) ~ \sigma ^{-1} ~ (physical)]
\endequation
\equation
\sigma ~ (symmetric) ~ = ~ \sigma ~ (physical) ~ - ~ b ~ p^2
\endequation

As in II, we take $p_{\alpha }$ for $p_{\alpha } ~ (physical)$, $\sigma$ for $\sigma ~ (physical)$, and use the hamiltonian equation in the symmetric reference system expressed, in the physical system, by the relation:

\equation
p_{0}^2 ~ = ~ p^2 ~ (1 ~ - ~ b ~ p^2 ~ \sigma ^{-1}) ~ + ~ (\sigma ~ - ~ b ~ p^2)^2 
\endequation
\noindent

If, in the symmetric system, we identify the constant value of $\sigma $ with the mass of the particle, we get in the physical system:

\equation
\sigma ~ ~ = ~ ~ \sigma _0 ~ ~ + ~ ~ b ~ ~ p^2
\endequation
\noindent
where $\sigma _0$ = = $T^{-1}$ $m$ , $m$ being the mass. When applied to (9), equation (10) gives:

\equation
p_{0}^2 ~ ~ = ~ ~ p^2 ~ ~ + ~ ~ \sigma _0^2~ ~ - ~ ~ b ~ p^4 ~ (\sigma _0 ~ ~ + ~ ~ b ~ ~ p^2)^{-1}
\endequation
\noindent
where, taking $b$ to be very small, we get the main QDRK model considered in our previous papers. Our model generalizes the standard Kaluza-Klein theory to a situation with new interactions producing a DRK scenario in the physical world. Following the method proposed in II, the $U$ operator used can be made unitary. To do this, we write:

\equation
p^2 ~ \partial /\partial p^2 ~ ~ =~ ~ ~ 1/2 ~ ~ p ~ \partial /\partial p ~ ~ = ~ ~ 1/2 ~ ~ D_{space}
\endequation
\noindent
where $D_{space} ~ = ~ ~ \Sigma _{i=1} ^3 ~ ~ p_i ~ \partial /\partial p_i$ . From (5) and (12) we get:

\equation
U' ~ ~ = ~ ~ exp ~ [ln ~ (1 ~ - ~ b ~ p^2 ~ \sigma ^{-1}) ~ ~(D_{space-2,\sigma } ~ - ~5/4)]
\endequation
\noindent

Our previous analysis of DRK (see papers I and II, and references therein) indicates that $b$ must not be a universal parameter but an operator depending on the properties of the object under consideration. For large bodies, we must have: $b ~ \simeq ~ b_0 ~ \sigma _0^{-1}$ where $b_0$ is a universal constant proportional to $T_0^2 ~ a^{-2}$ (see II).

~ 

A version of LDRK can also be obtained by writing instead of (5):

\equation
U (p_0 ~ , ~ p_i ~,~ \sigma  ~ , ~ T ~ , ~ T_0 ~ , ~ a) ~ = exp ~ [ln ~ (1 ~ + ~ b^L ~ p ~ \sigma ^{-1}) ~ ~D_{space-2 , \sigma-2}]
\endequation
\noindent
where $b^L$ is a constant and $D_{space-2 , \sigma-2} ~ = p^2 ~ \partial /\partial p^2 ~ + ~ \sigma ^2 ~ \partial /\partial \sigma ^2 $ another modified dilatation generator for the space and $\sigma$ momentum dimensions.. This leads to the equations:

\equation
p^2 ~ (symmetric) ~ = ~ p^2 ~ (physical) ~ [1 ~ - ~ b^L ~ p ~ (physical) ~ \sigma ^{-1} ~ (physical)]
\endequation
\equation
\sigma ~ (symmetric) ~ = ~ \sigma _0 ~ = ~ \sigma ~ (physical) ~ - ~ b^L ~ p ~ (physical)
\endequation
\noindent
where, as before, $\sigma _0 $ is a constant and, with the choice: $\sigma _0 $ = $T^{-1}$ = $m$ (mass of the particle), we get in terms of "physical" variables:

\equation
p_0^2~ ~ = ~ ~ p^2 ~ ~ + ~ ~ m^2 ~ ~ - ~ ~ b^L ~ p^3 ~ (m ~ +~ b^L ~ p)^{-1}
\endequation
\equation
\sigma ~ ~ = ~ ~ m ~ + ~ b^L ~ p
\endequation

For large bodies, $b^L$ must be a universal constant proportional to $T_0 ~ a^{-1}$ . The corresponding unitary operator is:

\equation
U' ~ ~ = ~ ~ exp ~ [ln ~ (1 ~ - ~ b^L ~ p ~ \sigma ^ {-1}) ~ ~(D_{space-2,\sigma-2 } ~ - ~1]
\endequation
\noindent

~ 

In equations (11) and (17), an important change occurs when $p$ becomes large enough so that $ b ~ p^2$ or $b^L ~ p$ is much larger than $m$. Then, both equations tend to a limit given by the equation $E^2 ~ = ~m^2$ . In the case of equation (11), the maximum of energy is attained in the region $p^2 ~ \sim ~ m ~ b^{-1}$ . If the particle under consideration is a proton, the maximum attainable energy can be at $\sim ~ ~ 10^{28} ~ eV$ (the Planck energy) if $b^{-1} ~ \sim ~ 10^{47} ~ eV$ , which means that $p^4 ~ \sim ~ b^{-1} ~ m^3$ at $p ~ \sim ~ 3 ~ . ~ 10^{18} ~ eV$ . This is an acceptable value from a phenomenological point of view, but lower values of $b$ are also permitted (see I and references therein). QDRK is thus perfectly compatible with DSR. The same analysis for LDRK leads to $ (b^L) ~ \sim ~ 10^{-19}$ for a proton, and in this case to $p^3 ~ \sim ~ (b^L)^{-1} ~ m^3$ for $p ~ \sim ~ 2 ~ . ~ 10^{15} ~ eV$ . This value seems too low for viable phenomenology, as its implications should already have been observed. Phenomenological problems of LDRK were discussed in I and in (Gonzalez-Mestres, 2000a).

~
~ 

\section{A deformed parton picture}
\label{DPP.sec}

~

With the conventional version of DRK, the parton model does not hold at energies where the deformation term becomes larger than the mass term. If the mass can be considered as a rest energy shared by the partons, the deformation term cannot be dealt with in the same way if its coefficient is a constant depending on the nature of the particle and the partons obey the same kinematical laws as if they where free particles. A possible way out was suggested in II, assuming that the coefficient of the deformation term is related to a new momentum dimension. If, in (11), $b$ has the properties of an inverse momentum, $b$ = $h$ $\eta ^{-1}$ , where $h$ is a dimensionless constant and $\eta$ the new momentum coordinate, we can assume $\eta $ to be shared among the partons. This makes the parton picture consistent. More precisely, we can write in the SRS :

\equation
p_{0}^2 ~ (symmetric) ~ = ~ p^2 ~ (symmetric) ~ + ~ \sigma ^2 ~ (symmetric)~ + ~ \eta ^2 ~ (symmetric)
\endequation
\noindent
and, using the transformation: 

\equation
U (p_0 ~ , ~ p_i ~,~ \sigma  ~ , ~ T ~ , ~ T_0 ~ , ~ a) ~ = exp ~ [ln ~ (1 ~ - ~ b^{\eta } ~ p^2 ~ \eta ^{-2}) ~ ~D_{space-2 , \eta -2}]
\endequation
\noindent
where $b^{\eta } ~ \simeq ~ h$ is a constant and, as before for $\sigma $ , $D_{space-2 , \eta -2} ~ = p^2 ~ \partial /\partial p^2 ~ + ~ \eta ^2 ~ \partial /\partial \eta ^2 $ is a new modified dilatation generator for the space and $\eta $ momentum dimensions, get:

\equation
p_0 ~ (symmetric) ~ = ~ p_0 ~ (physical)
\endequation
\equation
p^2 ~ (symmetric) ~ = ~ p^2 ~ (physical) ~ [1 ~ - ~ b^{\eta } ~ p^2 (physical) ~ \eta ^{-2} ~ (physical)]
\endequation
\equation
\sigma ~ (symmetric) ~ = ~ \sigma ~ (physical)
\endequation
\equation
\eta ^2 ~ (symmetric) ~ = ~ \eta ^2 ~ (physical) ) ~ - ~ b^{\eta } ~ p^2 (physical)
\endequation
\noindent
and, in the PRS :

\equation
p_{0}^2 ~ ~ = ~ ~ p^2 ~ ~ ~- ~ ~ b^{\eta } ~ ~ p^4 ~ \eta ^{-2} ~ ~ + ~ ~ \sigma ^2 ~ ~ + ~ ~ \eta ^2 ~ ~ - ~ b^{\eta } ~ ~ p^2
\endequation
\noindent
and, taking in the SRS the squared mass of the particle to be given by the expression:

\equation
m^2 ~ ~ = ~ ~ \sigma _0^2 ~ + ~ \eta _0^2
\endequation
\noindent
where $\sigma _0 $ and $\eta _0$ are the constant values of $\sigma $ and $\eta $ in the "symmetric" system, we get in the PRS :

\equation
p_{0}^2 ~ ~ = ~ ~ p^2 ~ ~ ~- ~ ~ b^{\eta } ~ ~ p^4 ~ ~ (\eta_0 ^2~ + ~ b^{\eta }~ p^2)^{-1} ~ ~ + ~ ~ m^2
\endequation
\equation
\eta ^2 ~ ~ = ~ \eta_0 ^2~ + ~ b^{\eta }~ p^2 
\endequation
\noindent

A consistent parton picture is then obtained if the partons share simultaneously 
all the components of the six-dimensional momentum: $p_0$ , $p_i $ , $\sigma $ and $\eta $ .

~ 

Taking $\sigma _0$ = 0 , $m^2 ~ = ~ \eta _0^2$ and $ b^{\eta } ~ \simeq ~ (a ~T_0^{-1})^2 ~ \ll ~ 1$ , we naturally get a consistent additive QDRK for large bodies, the deformation term being proportional to $p^2 ~ (p ~ m^{-1})^2$ and $ p ~ m^{-1}$ being a function of the velocity (see I , II and references therein). The same result is obtained if $\sigma _0$ is not zero but does not rise like the mass of the object under consideration or if $\sigma _0 ~ \eta 0^{-1}$ tends to a universal value for large bodies. 

~ 

If, instead, we take $b^{\eta } ~ \simeq ~ \rho ~ (a ~ \eta _0)^2$ , where $\rho $ is a universal constant, we get models of the kind:

\equation
p_{0}^2 ~ ~ = ~ ~ p^2 ~ ~ ~- ~ ~ \rho ~ a^2 ~ p^4 ~ ~ (1 ~ + ~ \rho ~ a^2~ p^2)^{-1} ~ ~ + ~ ~ m^2
\endequation
\noindent
where the coefficient of the deformation term is universal and the values of $\sigma _0$ and $\eta _0$ are restricted only by equation (27). Such models can be well suited in the elementary particle domain and $\eta _0$ can be very small (smaller than the photon mass), the mass of the elementary particle being basically given by $\sigma _0$ .

~ 

In a parton picture, the value of $\eta _0$ as a momentum integration constant in (28) is then to be multiplied by the fraction of momentum carried by the parton, but not in the expression $b^{\eta } ~ \simeq ~ \rho ~ (a ~ \eta _0)^2$ where $\eta _0$ plays the role of a dynamical constant having the same value for the "elementary" particle (e.g. a nucleon) and for its constituents (e.g. quarks and gluons), leading to the equation:.

\equation
p_{0}^2 ~ ~ = ~ ~ p^2 ~ ~ ~- ~ ~ \rho ~ a^2 ~ F^{-2} ~ p^4 ~ ~ (1 ~ + ~ \rho ~ a^2 ~ F^{-2} ~ p^2)^{-1} ~ ~ + ~ ~ m^2
\endequation
\noindent
where $F$ stands for the fraction of the total momentum, and the composite object would obey a DRK with the same value of $h$ as the partons.

~ 

More generally, we can extend DRK to a larger family of models by writing for large bodies:

\equation
b^{\eta } ~ \simeq ~ (a ~T_0^{-1})^{\lambda }
\endequation
\noindent
and, for elementary particles:

\equation
b^{\eta } ~ \simeq ~ \rho ~ a^{\lambda } ~ T_0^{2 ~ - ~ \lambda } ~ \eta _0^2
\endequation
\noindent
where the exponent $\lambda $ can be set independently of that of $p$ in the deformation term, $\lambda $ being a real number between 0 and 2 to be determined on phenomenological grounds. Although it may seem natural that $\lambda $ have the same value in (32) and (33), there is no compelling reason for this to be the case. We therefore can, even more generally, write:

\equation
b^{\eta } ~ \simeq ~ (a ~T_0^{-1})^{\lambda_1 }
\endequation
\noindent
for large bodies and:

\equation
b^{\eta } ~ \simeq ~ a^{\lambda_2 } ~ T_0^{2 ~ - ~ \lambda_2 } ~ \eta _0^2
\endequation
\noindent
for elementary particles.

~ 
~ 

\section{The Kirzhnits-Chechin model}
\label{kiche.sec}

~ 

At this stage, it seems necessary to remind and discuss in detail the Kirzhnits-Chechin model, as it was the first to consider a "physical" reference system and a "symmetric" one. Even if it turns out that the model cannot be used such as it was formulated, it turns out that our more recent proposals since 1997, combined with the Magueijo-Smolin transformation, allow for consistent generalizations of the model leading to viable phenomenology and to useful theoretical concepts.

~ 

Kirzhnits and Chechin suggested long ago (19971 and 19972) that the absence of Greisen-Zatsepin-Kuzmin (GZK) cutoff could be explained by models where special relativity was replaced by a new formulation based on the Finsler space, where the usual relation $ p_{0}^2 ~ = ~ p^2 ~ + ~ m ^2$ being replaced by:

\equation
f ~ (p_{\alpha}) ~ (p_{0}^2 ~ - ~ p^2) ~ = ~ m ^2
\endequation
\noindent
where $f ~ (p_{\alpha})$ is a homogeneous positive function of the four momenta of zero degree. These authors used:

\equation
f ~ (p_{\alpha}) ~ ~ = ~ ~ f ~ (\xi )
\endequation
\noindent
where:

\equation
\xi ~ = ~ [p^2 ~ (p_{0}^2 ~ - ~ p^2)^{-1}]
\endequation
\noindent
$f ~ (0) ~ = ~ 1$ and $f$ was supposed to tend to some constant $f ~ (\infty )$ in the range 0.01 - 0.1 as $\xi ~ \rightarrow ~ \infty $ . This amounts to a shift of the effective squared mass by a factor 10 to 100 above some critical value of $\xi $ . The dispersion relation for the photon was assumed to have no deformation, and $f$ was taken to be $f ~ (p_{\alpha}) ~ = ~ f ~ (\infty )$ for this particle. For a massive particle, it is assumed that $f$ can be expanded as: 

\equation
f ~ (\xi ) ~ ~ \simeq ~ ~ 1 ~ - ~ \alpha ~ \xi ^2 ~ + ~ ...
\endequation
\noindent
leading, to a first sight, to models close to QDRK. For the proton, the term $\alpha ~ \xi ^2 $ becomes $\approx $ 1 at $E_p$ $\approx $ $10^{20} ~ eV$ if $\alpha $ $\approx $ $10^{-44}$ .

~ 

Noticing that the four-momentum $p_{\alpha}^*$ defined as:

\equation
p_{\alpha}^* ~ ~ = ~ ~ [f ~ (\xi )]^{1/2}
 ~ p_{\alpha}
\endequation
\noindent
transforms like a four-vector under Lorentz transformations, Kirzhnits and Chechin considered central collisions of ultra-high energy protons with microwave background radiation photons. The thermal spectrum, usually described in the laboratory frame (supposed to be close to the rest frame suggested by the microwave background radiation spectrum) by the Planck distribution $H ~ = ~ exp ~ [- ~ E_{\gamma } ~ (k ~ T_B)^{-1}]$ where $E_{\gamma }$ is the photon energy, $T_B$ the microwave background radiation temperature and $k$ the Boltzmann constant, becomes in the rest frame of the incoming proton a modified distribution $H_p$ given by:

\equation
ln ~ H_p ~ = ~ - ~ E_{\gamma } ~ m ~ [2 ~ E_p ~ k ~ T_B]^{-1} ~ [f ~ (\xi _p)]^{-1/2}
\endequation
\noindent
$\xi _p$ being the value of $\xi $ for the incoming proton in the laboratory frame and $m$ the proton mass. Because of the factor $[f ~ (\xi _p)]^{1/2}$ , the photon energies are smaller than expected and pion photoproduction via the $\Delta $ resonance is inhibited, as the resonance can no longer be formed. A similar argument can be formulated in the laboratory rest frame, noticing the shift by a factor 10 to 100 of the effective squared masses of pions and nucleons.

~ 

The Kirzhnits-Chechin (Kch) model can be incorporated into the operator formalism by writing:

\equation
2 ~ ln ~ U (p_0 ~ , ~ p_i ~,~ \sigma  ~ , ~ T ~ , ~ T_0 ~ , ~ a) ~ ~ = ~ ~ ln ~ f ~ (\xi )~ ~ D_{space-time}
\endequation
\noindent
where $ D_{space-time}$ = $\Sigma _{\alpha } ~ p_{\alpha} ~ \partial /\partial p_\alpha $ and $p_{\alpha } ~ (symmetric)$ is to be identified with $ p_{\alpha}^*$ .The corresponding unitary operator is:

\equation
2 ~ ln ~ U' ~ (p_0 ~ , ~ p_i ~,~ \sigma ~ , ~ T ~ , ~ T_0 ~ , ~ a) ~ ~ = ~ ~ ln ~ f ~ (\xi )~ ~ (D_{space-time} ~ - ~ 2)
\endequation
\noindent
and, for large bodies, it agrees with our proposal (see I and II) that the coefficient of the quadratic deformation be proportional to $m^{-2}$ in order to guarantee the consistency of the kinematics. Like our models, the KCh model assumes the $p_{\alpha}$'s to be conserved and additive for free particles. But, contrary to our models, it leads to the same property for elementary particles. It was actually with this assumption that Kirzhnits and Chechin proposed their model to be a solution to the puzzle generated by the experimental absence of GZK cutoff. However, it seems to us that the KCh model so defined may suffer from some pathologies. Actually, these authors did not completely write down an explicit example of their model, so that its consistency has never been proven.

~ 

A typical formulation of the KCh model can be:

\equation
f ~ (\xi ) ~ ~ = ~ ~ (1 ~ + ~ \alpha_1 ~ \xi ^2) ~ (1 ~ + ~ \alpha_2 ~ \xi ^2)^{-1}
\endequation
\noindent
where $\alpha _2 $ $\approx $ $10^{-44}$ and $\alpha _1$ $\approx $ $10^{-45}$ to $10^{-46}$ , so that, as conjectured by Kirzhnits and Chechin, $f ~ (\infty )$ is in the range 0.01 - 0.1 . We then have:

\equation
(1 ~ + ~ \alpha_1 ~ \xi ^2) ~ (p_{0}^2 ~ - ~ p^2) ~ = ~ m ^2 ~ (1 ~ + ~ \alpha_2 ~ \xi ^2)
\endequation
\noindent
which leads to the equation:

\equation
[(p_0^2 ~ - ~ p^2)^2 ~ + ~ \alpha_1 ~p^4] ~ (p_{0}^2 ~ - ~ p^2) ~ = ~ m ^2 ~ [(p_{0}^2 ~ - ~ p^2)^2 ~ + ~ \alpha_2 ~ p^4]
\endequation
\noindent
and, with a trivial approximation, to:

\equation
[(p_0 ~ - ~ p)^2 ~ + ~ \alpha_1 ~p^2/4] ~ (p_{0} ~ - ~ p) ~ \simeq ~ m ^2 ~ (2p)^{-1} ~ [(p_{0} ~ - ~ p)^2 ~ + ~ \alpha_2 ~ p^2/4]
\endequation
\noindent
that can be solved by standard methods to obtain $p_0$ as a function of $p$ . Another way to handle the equation would be to write:

\equation
\epsilon ~ ~ = ~ ~ 2 ~ (p_0 ~ - ~ p) ~ m^{-2} ~ p
\endequation
\equation
[\epsilon ^2 ~ m^4 ~ + ~ \alpha_1  p^4] ~ \epsilon ~ \simeq ~ \epsilon ^2 ~ m^4 ~ + ~ \alpha_2 ~p^4
\endequation
\noindent
leading to :
\equation
\alpha_2 ~(p/m)^4 ~ \simeq ~ \epsilon ^2 ~ (\epsilon ~ - ~ 1) ~ (1 ~ - ~ \epsilon ~ \alpha_1 ~ \alpha_2^{-1})^{-1}
\endequation
\noindent
and use this equation to illustrate with a few numerical values the instability of the ultra-high energy proton in the conditions of the KCh model. Taking $\alpha_1 ~ \alpha_2^{-1}$ = 0.1 and comparing the values of $p$ for $\epsilon ~ = ~ 2$ and $\epsilon ~ = ~ 2.5$ , we readily get: 

\equation
\alpha_2 ~(p/m)^4 ~ (\epsilon ~ = ~ 1.43) ~ \simeq ~ 1.02
\endequation
\equation
\alpha_2 ~(p/m)^4 ~ (\epsilon ~ = ~ 2) ~ \simeq ~ 5
\endequation
\equation
\alpha_2 ~(p/m)^4 ~ (\epsilon ~ = ~ 2.5) ~ \simeq ~ 12.5 
\endequation
\equation
\alpha_2 ~(p/m)^4 ~ (\epsilon ~ = ~ 5) ~ \simeq ~ 200 
\endequation
\equation
\alpha_2 ~(p/m)^4 ~ (\epsilon ~ = ~ 8) ~ \simeq ~ 2240 
\endequation
\equation
\alpha_2 ~(p/m)^4 ~ (\epsilon ~ = ~ 9.5) ~ \simeq ~ 15342.5 
\endequation
\noindent
so that:

~ 

- The ratio $[\alpha_2 ~(p/m)^4 ~ (\epsilon ~ = ~ 2)]^{1/4} ~ [\alpha_2 ~(p/m)^4 ~ (\epsilon ~ = ~ 1.43)]^{-1/4}$ = 1.488 differs only by $6\% $ from the ratio between the two values of $\epsilon $, and the difference between two values of $p_0 ~ - ~ p$ will be of the same order. But the two momenta differ by $50\%$ . Absorbing a $\simeq ~ 4 ~. ~ 10^{-4} ~ eV$ microwave background photon in the laboratory system will allow the $\simeq $ 1.5 . $10^{11}$ $m$ proton to emit a $\simeq ~ 0.5 ~ . ~ 10^{20} ~ eV$ photon. 

~ 

- Again, the ratio $[\alpha_2 ~(p/m)^4 ~ (\epsilon ~ = ~ 2.5)]^{1/4} ~ [\alpha_2 ~(p/m)^4 ~ (\epsilon ~ = ~ 2)]^{-1/4}$ = 1.2574 differs only by $6\% $ from the ratio between the two values of $\epsilon $, and so will the two values of $p_0 ~ - ~ p$ . The two momenta differ by $26\%$ . Absorbing a $\simeq ~ 8 ~. ~ 10^{-4} ~ eV$ microwave background photon in the laboratory system will be enough for the $\simeq $ 1.9 . $10^{11}$ $m$ proton to emit a $\simeq ~ 0.3 ~ . ~ 10^{20} ~ eV$ photon. 

~ 

- For $\epsilon $ = 5 , we get $[\alpha_2 ~(p/m)^4 ~ (\epsilon ~ = ~ 5)]^{1/4} ~ [\alpha_2 ~(p/m)^4 ~ (\epsilon ~ = ~ 2)]^{-1/4}$ = 2.06 , so that the value of $p_0 ~ - ~ p$ will be 2.5 times higher for $\epsilon $ = 5 , as compared to $\epsilon $ = 2 , and a 3 . $10^{11}$ $m$ proton can spontaneoulsy emit a $\simeq ~ 1.4 ~ . ~ 10^{20} ~ eV$ photon. 

~ 

- For $\epsilon $ = 8 , we get $[\alpha_2 ~(p/m)^4 ~ (\epsilon ~ = ~ 8)]^{1/4} ~ [\alpha_2 ~(p/m)^4 ~ (\epsilon ~ = ~ 2)]^{-1/4}$ = 4.6 , so that the value of $p_0 ~ - ~ p$ will be 1.74 times higher for $\epsilon $ = 8 , as compared to $\epsilon $ = 2 , and a 7 . $10^{11}$ $m$ proton can spontaneoulsy emit a $\simeq ~ 5.5 ~ . ~ 10^{20} ~ eV$ photon. 

~ 

- For $\epsilon $ = 9.5 , we get $[\alpha_2 ~(p/m)^4 ~ (\epsilon ~ = ~ 8)]^{1/4} ~ [\alpha_2 ~(p/m)^4 ~ (\epsilon ~ = ~ 2)]^{-1/4}$ = 7.44 , so that the value of $p_0 ~ - ~ p$ will be 1.28 times higher for $\epsilon $ = 9.5 , as compared to $\epsilon $ = 2 , and a 8.9 . $10^{11}$ $m$ proton can spontaneoulsy emit a $\simeq ~ 7.4 ~ . ~ 10^{20} ~ eV$ photon. 

~ 

In all these cases, pion emission, instead of photon emission, is also possible and, not only the GZK cutoff is at work, but the ultra-high energy proton becomes unstable and decays to a lower energy proton. The situation would be even worse taking $\alpha_1 ~ \alpha_2^{-1}$ = 0.01 . We therefore conclude that such a model would be of no practical use in the UHECR region. Thus, the question is how general are these diseases of the KCh model. To answer this question, we notice that:

\equation 
\epsilon ~ ~ = ~ ~ [f ~ (\xi )]^{-1/2}
\endequation
\noindent
so that $p^*_1$ $<$ $p^*_2$ for $p_1$ $>$ $p_2$ if $\epsilon ~ (p_1) ~ [\epsilon ~ (p_2)]^{-1}$ $>$ $p_1 ~ p_2^{-1}$ . If such a situation occurs, it means that in the region under consideration there are at least two possible values of $p$ for a given value of $p^*$, and that the highest value of $p$ corresponds to an unstable state in the "physical" reference system. To avoid such an instability, we must require that:

\equation 
d ~ ln ~ [f ~ (\xi )] ~ / ~ d ~ ln ~ p ~ ~ > ~ ~ - ~ 2
\endequation
\noindent
so that $[f ~ (\xi )]^{-1/2}$ cannot grow faster than $p$. However, if this is the case, the factor $ E_p ~ [f ~ (\xi p)]^{1/2}$ in (59) has a lower bound and the GZK mechanism cannot be inhibited at ultra-high energies. A similar criticism can be opposed to the phenomenological analysis presented by Chechin and Vavilov, 1999 , which directly uses asymptotic values for the $f$ function considered. From the form of the condition (59), it follows that the problem raised is quite general and independent of the asymptotic value of $f$ at large $\xi $ .

~ 

However, the models suggested in II and in the present paper can be regarded as generalizations of the KCh model in the sense that, with the Magueijo-Smolin transformation, modified dilatation operators can be introduced multiplied by functions of variables which are scalars with respect to these dilatations. Thus, the Finsler space idea is generalized to a much larger family of models and spaces, including situations where a preferred inertial frame exists.

~ 
~ 

\section{The Sato-Tati model}
\label{satotati.sec}

~ 

In 1971, Sato et Tati (1972) proposed, as a solution to the UHECR puzzle, the existence of a preferred reference frame and the impossibility for hadronic matter to exist above a value of the Lorentz factor $\simeq ~ 10^{11}$ with respect to this frame. This model involves a very strong dynamical assumption on the production of hadronic matter at ultra-high energy that is not present in our models. Rather than with the structure of space-time, this model seems to be concern with the dynamical properties of vacuum in our Universe and of hadronic matter. Even with a privileged vacuum rest frame, the Sato-Tati model (ST) can perfectly well incorporate exact relativistic kinematics and have only a sharp dynamical threshold for the inhibition of hadronic particle production.

~ 

Sato and Tati assume that there is a maximum allowed value of $E/m$ , $\gamma _ {max}$ , for hadronic matter and that, even if the cutoff is not sharp and may have a long tail, it is enough to sharply reduce the production probability of hadrons above $\gamma _ {max}$ . However, it is not obvious why $2. 10^{19} ~ eV $ pion production would be suppressed when a $E ~ \simeq ~ 10^{20} ~ eV$ proton hits microwave background radiation, but the production of $3 ~ 10^{20} ~ eV$ protons would not be inhibited in acceleration processes or decays of ultra-heavy particles.

~ 

Furthermore, the suppression itself is unclear, as $E ~ \simeq ~ 10^{20} ~ eV$ would in any case be an acceptable energy for a $\Delta $ resonance. The $m^2 ~ (2 ~ p)^{-1}$ mass term for a $E ~ \simeq ~ 10^{20} ~ eV$ proton is $\simeq $ 8.8 times larger than its equivalent for a $2. 10^{19} ~ eV $ pion  and $\simeq $ 4.4 times larger if the pion energy is $10^{19} ~ eV $ ($m^2 ~ (2 ~ p)^{-1} $ $\simeq $ $10^{-3} ~ eV$). And, even if the emission of a neutral pion is forbidden by the $\gamma _ {max}$ cutoff, this has no reason to be the case for its two-photon decay product. In most scenarios, it seems possible to preserve the GZK cutoff if the Sato-Tati cutoff on hadronic matter does not affect $10^{19} ~ eV $ pions or photons. 

~ 

However, the question of whether hadronic matter can exist above some critical value of $E/m$ in the vacuum rest frame is a fundamental one. Some aspects of this problem will be discussed in a forthcoming paper. 

~ 
~ 

\section{Conclusion and comments}
\label{concl.sec}

~

The method presented here can be made quite general by introducing new space-time dimensions as well as a large family of generalized dilatation operators, and using in particular, in the $U$ operators, expressions which are scalars with respect to these generalized dilatations. If necessary, a vacuum rest frame can also be introduced together with a set of $U$ operators depending explicitly on the speed of the particle with respect to this frame. 

~ 

With this technique, QDRK and LDRK models can be produced, Dirac equations can be deformed in a similar way and calculations are rather easy. The new approach provides a quite large extension of the previous KCh model based on Finsler algebras and, although the original ansatz by Kirzhnits and Chechin does not seem to work, the kind of generalization proposed here and in II allows to include in the pattern the DRK models we developed since 1997 and which seem well-suited for phenomenology.

~ 

Thus, the Magueijo-Smolin technique turns out to be extremely useful to discuss patterns of Lorentz symmetry violation. Properties of space-time are also a crucial item, and a complete study is in progress in view of phenomenological applications.

~ 
~ 

%
%
%
\vspace{1ex}
\begin{center}
{\bf References}
\end{center}
%
\noindent
Amelino-Camelia, G., 2001, Phys.Lett. B510, 255. \\
Amelino-Camelia, G., 2002, Int. J. Mod. Phys. D11, 35.\\
Amelino-Camelia, G., Benedetti, D. and D'Andrea, F., 2002, paper hep-th/0201045.\\
Bruno, R., Amelino-Camelia, G. and Kowalski-Gilman, J., 2001, Phys. Lett. B522, 133.\\
Chechin, V.A. and Vavilov, Yu.N.,1999, Proceedings of the ICRC 1999 Conference, paper HE.2.3.07, in http://krusty.physics.utah.edu/$\sim $icrc1999 .\\
Gonzalez-Mestres, L., 1997a, paper nucl-th/9708028.\\
Gonzalez-Mestres, L., 1997b, Proc. of the International
Conference on Relativistic Physics and some of its Applications, Athens June 1997, paper physics/9709006. \\
Gonzalez-Mestres, L., 1997c, paper physics/9704017.\\
Gonzalez-Mestres, L., 1997d, Proc. 25th ICRC, Vol. 6, p. 113, physics/9705031.\\
Gonzalez-Mestres, L., 1997e, paper physics 9706032.\\
Gonzalez-Mestres, L., 1998, Proc. COSMO 97, Ambleside September 1997,
World Scient., p. 568, paper physics/9712056. \\
Gonzalez-Mestres, L., 2000a, paper physics/003080 , referred to as I.\\
Gonzalez-Mestres, L., 2000b, paper astro-ph/0011181, Proceedings of the International Symposium on High Energy Gamma-Ray Astronomy, Heidelberg, Germany, June 26-30, 2000. \\
Gonzalez-Mestres, L., 2000c, paper astro-ph/0011182, same Proceedings.\\
Gonzalez-Mestres, L., 2002, paper hep-th/0208064, referred to as II.\\
Kirzhnits, D.A. and Chechin, V.A., 1971, ZhETF Pis. Red. 4, 261.\\
Kirzhnits, D.A. and Chechin, V.A., 1972, Yad. Fiz. 15, 1051. \\ 
Lukierski, J. and Nowicki, A., 2002, paper hep-th/0203065 , and references therein. \\
Magueijo, J. and Smolin, L., 2001, paper hep-th/0112090. \\
Magueijo, J. and Smolin, L., 2002, paper gr-qc/0207085. \\
Poincar\'e, H., 1895, "A propos de la th\'eorie de M. Larmor",
L'Eclairage \'electrique, Vol. 5,  5.\\
Poincar\'e, H., 1901, "Electricit\'e et Optique: La lumi\`ere
et les th\'eories \'electriques", Ed. Gauthier-Villars, Paris.\\
Poincar\'e, H., 1905, "Sur la dynamique de l'\'electron", C.
Rend. Acad. Sci. Vol. 140, p. 1504.\\
Sato, H. and Tati, T. , 1972, Progr. Theor. Phys. 47, 1788.\\
\end{document}